\documentclass[12pt]{article}
\usepackage{amssymb}
\usepackage{amsmath}
\usepackage{amsfonts}

\newtheorem{theorem}{Theorem}

\def\ket #1{\vert #1\rangle}
\def\bra #1{\langle #1\vert}
\def\braket #1#2{\langle #1 \vert #2\rangle}

\begin{document}

\title{On complementary channels and the additivity problem}
\author{A. S. Holevo}
\date{}
\maketitle

\begin{abstract}
We explore complementarity between output and environment of a quantum
channel (or, more generally, CP map), making an observation that the output
purity characteristics for complementary CP maps coincide. Hence, validity
of the mutiplicativity/additivity conjecture for a class of CP maps implies
its validity for complementary maps. The class of CP maps complementary to
entanglement-breaking ones is described and is shown to contain diagonal CP
maps as a proper subclass, resulting in new class of CP maps (channels) for
which the multiplicativity/additivity holds. Covariant and Gaussian channels
are discussed briefly in this context.
\end{abstract}

\bigskip In what follows $\mathcal{H}_{A},\mathcal{H}_{B},\dots $ will
denote (finite dimensional) Hilbert spaces of quantum systems $A,B,\dots .
\mathfrak{M}\left( \mathcal{H}\right) $ denotes the algebra of all operators, $
\mathfrak{S}\left( \mathcal{H}\right) --$ the convex set of density operators
(states) and $\mathfrak{P}\left( \mathcal{H}\right) =\mathrm{ext}
\mathfrak{S}\left( \mathcal{H}\right) -$ the set of pure states (one-dimensional
projections) in $\mathcal{H}.$ For a natural $d,$ $\mathcal{ \ \ \ \ H}_{d}$
denotes the Hilbert space of $d-$dimensional complex vectors, and
$\mathfrak{M}_{d}$ -- the algebra of all complex $d\times d-$ matrices.

\bigskip Given three finite spaces $\mathcal{H}_{A},\mathcal{H}_{B},\mathcal{
H}_{C}$ and a linear operator $V:\mathcal{H}_{A}\rightarrow \mathcal{H} _{B}\otimes
\mathcal{H}_{C}$, the relation
\begin{equation}
\Phi _{B}(\rho )=\mathrm{Tr}_{\mathcal{H}_{C}}V\rho V^{\ast },\quad \Phi
_{C}(\rho )=\mathrm{Tr}_{\mathcal{H}_{B}}V\rho V^{\ast };\quad \rho \in
\mathfrak{M}\left( \mathcal{H}_{A}\right)  \label{compl}
\end{equation}
\bigskip defines two CP maps $\Phi _{B}:\mathfrak{M}\left( \mathcal{H}
_{A}\right) \rightarrow \mathfrak{M}\left( \mathcal{H}_{B}\right) ,$ $\Phi
_{C}:\mathfrak{M}\left( \mathcal{H}_{A}\right) \rightarrow \mathfrak{M} \left(
\mathcal{H}_{C}\right) ,$ which will be called mutually \textit{ complementary}. If
$V$ is an isometry, both maps are trace preserving (TP) i.e. channels. The name
\textquotedblleft complementary channels\textquotedblright\ is taken from the paper
\cite{DS}, where they were used to define quantum version of degradable channels.

The Stinespring dilation theorem implies that for given a CP map (channel) a
complementary always exists. In the Appendix we give a proof which also
clarifies in what sense the complementary map is unique. It follows that for
a given CP map $\Phi _{B}$, any two channels $\Phi _{C},\Phi _{C^{\prime }}$
complementary to $\Phi _{B}$ are equivalent in the sense that there is a
partial isometry $W:\mathcal{H}_{C}\rightarrow \mathcal{H}_{C^{\prime }}$
such that
\begin{equation}
\Phi _{C^{\prime }}(\rho )=W\Phi _{C}(\rho )W^{\ast },\quad \Phi _{C}(\rho
)=W^{\ast }\Phi _{C^{\prime }}(\rho )W,  \label{equiv}
\end{equation}
for all $\rho .$ Dilations with the minimal dimensionality $d_{C}$ are
called \textit{minimal}. Any two minimal dilations are isometric (i.e. $W$
is an isometry from $\mathcal{H}_{C}$ onto $\mathcal{H}_{C^{\prime }}$). By
performing a Stinespring dilation for a complementary CP map one obtains a
map equivalent to the initial one in the sense (\ref{equiv}). Thus the
complementarity is a relation between the equivalence classes of CP maps.

To simplify formulas we shall also use the notation $\tilde{\Phi}$ for the map
which is complementary to $\Phi $.

Consider the following \textquotedblleft measures of output
purity\textquotedblright\ of a CP map $\Phi $
\begin{equation}
\nu _{p}(\Phi )=\max_{\rho \in \mathfrak{S}(\mathcal{H})}[\mathrm{Tr}\Phi
(\rho )^{p}]^{1/p},\quad 1\leq p,  \label{purity}
\end{equation}
introduced in \cite{AHW}. For $p=\infty $ one puts $\nu _{\infty }(\Phi
)=\max_{\rho \in \mathfrak{S}(\mathcal{\ H})}\left\Vert \Phi (\rho
)\right\Vert .$ In the case of channel $\Phi $, further useful
characteristics are the minimal output entropy
\begin{equation*}
\check{H}(\Phi )=\min_{\rho \in \mathfrak{S}(\mathcal{H})}H(\Phi (\rho )),
\end{equation*}
where $H(\sigma )=-\mathrm{Tr}\sigma \ln \sigma $ is the von Neumann entropy
of a density operator $\sigma ,$ and its \textit{convex closure}
\begin{equation*}
\hat{H}_{\Phi }\left( \rho \right) =\min_{\rho =\sum_{x}\pi (x)\rho
(x)}\sum_{x}\pi (x)H(\Phi (\rho (x))),
\end{equation*}
where the minimum is taken over all possible convex decompositions of the density
operator $\rho $ into pure states $\rho (x)\in \mathfrak{S}(\mathcal{ \ H})$
\cite{HS}. By convexity argument, all these quantities remain unchanged if we
replace $\mathfrak{S}(\mathcal{H})$ by $\mathfrak{P}( \mathcal{H})$ in their
definitions.

\begin{theorem}
If one of the relations
\begin{equation}
\nu _{p}\left( \Phi _{1}\otimes \Phi _{2}\right) =\nu _{p}\left( \Phi
_{1}\right) \nu _{p}\left( \Phi _{2}\right) ,  \label{n1p}
\end{equation}
\begin{equation}
\check{H}\left( \Phi _{1}\otimes \Phi _{2}\right) =\check{H}\left( \Phi
_{1}\right) +\check{H}\left( \Phi _{2}\right) ,  \label{maddi}
\end{equation}
\begin{equation}
\hat{H}_{\Phi _{1}\otimes \Phi _{2}}(\rho _{12})\geq \hat{H}_{\Phi
_{1}}(\rho _{1})+\hat{H}_{\Phi _{2}}(\rho _{2})  \label{hsadd}
\end{equation}
holds for the CP maps (channels) $\Phi _{1},\Phi _{2}$, then similar
relation holds for the pair of their complementary maps $\tilde{\Phi}_{1},
\tilde{\Phi}_{2}.$ If one of these relations holds for given $\Phi _{1}$ and
arbitrary $\Phi _{2},$ then similar relation holds for complementary $\tilde{
\Phi}_{1}$ and arbitrary $\Phi _{2}.$
\end{theorem}

\textbf{Remark.} Let us recall that for two given channels $\Phi _{1},\Phi _{2}$,
the property (\ref{n1p}) with $p\in \lbrack 1,1+\varepsilon ]$ implies
(\ref{maddi}) by differentiation \cite{AHW}. The property (\ref {hsadd}), which is
equivalent to the additivity of the $\chi -$capacity (the Holevo capacity) with
arbitrary input constraints \cite{HS}, implies both additivity of the $\chi -
$capacity and (\ref{maddi}) by the arguments similar to that for the
superadditivity of entanglement of formation, see e. g. \cite{S1}. On the other
hand, assuming that (\ref{n1p}) with $p\in \lbrack 1,1+\varepsilon ]$ holds for all
CP maps $\Phi _{1},\Phi _{2}$ implies (\ref{maddi}), (\ref{hsadd}) for all
channels, and these two properties, as well as additivity of the $\chi -$capacity,
are globally equivalent, i. e. if one holds for all channels, another holds for all
channels as well \cite{S1}.

\textit{Proof. }If $\rho =|\psi \rangle \langle \psi |$ for some $|\psi \rangle \in
\mathcal{H}_{A},$ then Hermitian operators $\Phi (\rho ),\tilde{ \Phi}(\rho )$ have
the same nonzero eigenvalues. Indeed, $\Phi (\rho ), \tilde{\Phi}(\rho )$ are
partial traces of the operator $|\psi _{BC}\rangle \langle \psi _{BC}|,$ where
$|\psi _{BC}\rangle =V|\psi \rangle \in \mathcal{ \ \ H}_{B}\otimes
\mathcal{H}_{C},$ then the proof goes in the same way as in the case of normalized
vectors (see, e.g. \cite{NC}, Theorem 2.7).

Both $\mathrm{Tr}\sigma ^{p}$ and $H(\sigma )$ are universal functions of nonzero
eigenvalues of a Hermitian operator $\sigma .$ From the definitions of $\nu _{p},
\check{H}$ and $\hat{H}$ it follows that for arbitrary CP map $ \Phi $
\begin{equation}
\nu _{p}(\tilde{\Phi})\,=\,\nu _{p}(\Phi ).  \label{nbc}
\end{equation}
Moreover, if $\Phi \ $is a channel, then
\begin{eqnarray}
\check{H}(\tilde{\Phi}) &=&\check{H}(\Phi ),\quad  \label{hbc} \\
\hat{H}(\tilde{\Phi}) &=&\hat{H}(\Phi ).  \label{hsbc}
\end{eqnarray}

Now notice that if $\Phi _{j},\tilde{\Phi}_{j},j=1,2,$ are two pairs of
complementary CP maps, then $\Phi _{1}\otimes \Phi _{2}$ and $\tilde{\Phi}
_{1}\otimes \tilde{\Phi}_{2}$ are complementary. For this take $\mathcal{H}
_{B}=\mathcal{H}_{B_{1}}\otimes \mathcal{H}_{B_{2}},\mathcal{H} _{C}=
\mathcal{H}_{C_{1}}\otimes \mathcal{H}_{C_{2}}$ and $V=V_{1}\otimes V_{2}.$
Summarizing all these facts, we get the statement. $\square $

Assume that a CP map $\Phi :\mathcal{M}(\mathcal{H})\rightarrow \mathcal{M}(
\mathcal{H}^{\prime })$ is given by Kraus representation
\begin{equation}
\Phi (\rho )=\sum_{\alpha =1}^{\tilde{d}}V_{\alpha }\rho V_{\alpha }^{\ast },
\label{kr}
\end{equation}
then a complementary map $\tilde{\Phi}:$ $\mathcal{M}(\mathcal{H} )\rightarrow
\mathcal{M}_{\tilde{d}}$ is given by
\begin{equation}
\tilde{\Phi}(\rho )=\left[ \mathrm{Tr}V_{\alpha }\rho V_{\beta }^{\ast } \right]
_{\alpha ,\beta =\overline{1,\tilde{d}}}=\left[ \mathrm{Tr}\rho V_{\beta }^{\ast
}V_{\alpha }\right] _{\alpha ,\beta =\overline{1,\tilde{d}} },  \label{cmpm}
\end{equation}
since $V=\sum_{\alpha =1}^{\tilde{d}}\oplus V_{\alpha }$ is a map from $
\mathcal{H}$ to $\sum_{\alpha =1}^{\tilde{d}}\oplus \mathcal{H}^{\prime }\simeq
\mathcal{H^{\prime }\otimes H}_{\tilde{d}}$ for which $\Phi ,\tilde{ \Phi}$ are
given by the partial traces (\ref{compl}), see \cite{H3}. By writing the trace in
$\mathcal{H}^{\prime }$ with respect to an orthonormal basis $\{e_{j}^{\prime }\}$,
we have the Kraus representation
\begin{equation}
\tilde{\Phi}(\rho )=\sum_{j=1}^{d^{\prime }}\tilde{V}_{j}\rho \tilde{V} _{j}^{\ast
},  \label{ckd}
\end{equation}
where $(\tilde{V}_{j})_{\alpha }=\bra{e'_j}V_{\alpha }.$ One can check by
direct computation that applying the same procedure to $\tilde{\Phi},$ one
obtains the map $\tilde{\tilde{\Phi}}$ which is isometric to $\Phi .$

A CP map $\Phi :\mathcal{M}(\mathcal{H})\rightarrow \mathcal{M}(\mathcal{H}
^{\prime })$ is \textit{entanglement-breaking} if it has a Kraus
representation with rank one operators $V_{\alpha }$ \cite{HRS}:
\begin{equation}
\Phi (\rho )=\sum_{\alpha =1}^{\tilde{d}}|\varphi _{\alpha }\rangle \langle
\psi _{\alpha }|\rho |\psi _{\alpha }\rangle \langle \varphi _{\alpha }|.
\label{eb}
\end{equation}
Such a CP map is channel if and only if the (over)completeness relation
\begin{equation*}
\sum_{\alpha =1}^{\tilde{d}}|\psi _{\alpha }\rangle \langle \varphi _{\alpha
}|\varphi _{\alpha }\rangle \langle \psi _{\alpha }|=I
\end{equation*}
is fulfilled. The complementary map $\tilde{\Phi}:\mathcal{M}(\mathcal{H}
)\rightarrow \mathcal{M}_{\tilde{d}}$ is
\begin{equation}
\tilde{\Phi}(\rho )=\left[ c_{\alpha \beta }\langle \psi _{\alpha}|\rho
|\psi _{\beta}\rangle \right] _{\alpha ,\beta =\overline{1,\tilde{d}}},
\label{gdiag}
\end{equation}
where $c_{\alpha \beta }=\langle \varphi _{\beta }|\varphi _{\alpha}\rangle . $
Notice that by the Kolmogorov decomposition, arbitrary nonnegative definite matrix
can be represented in such form. In the special case where $ \left\{ \psi _{\alpha
}\right\} _{\alpha =\overline{1,\tilde{d}}}$ is an orthonormal base in
$\mathcal{H}$, (\ref{gdiag}) is \textit{diagonal }CP map \cite{Kdiag}. Diagonal
channels, which are characterized by additional property $c_{\alpha \alpha }\equiv
1,$ were also earlier considered in \cite {DS} under the name of dephasing
channels. From (\ref{eb}) we see that the diagonal maps are complementary to a
particular class of entanglement-breaking maps, namely to c-q maps. For another
special subclass of entanglement-breaking maps, the q-c maps, $\left\{ \varphi
_{\alpha }\right\} _{\alpha =\overline{1,\tilde{d}}}$ is an orthonormal base in $
\mathcal{H},$ so that $c_{\alpha \beta }=\delta _{\alpha \beta },$ and the
complementary map is easily seen to be of the same subclass.

\bigskip Let us rewrite (\ref{gdiag}) in the form
\begin{equation*}
\tilde{\Phi}(\rho )=\sum_{\alpha ,\beta =1}^{\tilde{d}}c_{\alpha \beta
}|e_{\alpha }\rangle \langle \psi _{\alpha }|\rho |\psi _{\beta }\rangle
\langle e_{\beta }|
\end{equation*}
where $\left\{ e_{\alpha }\right\} $ is the canonical base for $\mathcal{H} _{
\tilde{d}}.$ Representing $c_{\alpha \beta }=\sum_{j=1}^{d^{\prime}}\bar{ v }
_{\beta j}v_{\alpha j}$ by Kolmogorov decomposition and denoting
\begin{equation}
\tilde{V}_{j}=\sum_{\alpha =1}^{\tilde{d}}v_{\alpha j}|e_{\alpha }\rangle
\langle \psi _{\alpha }|,  \label{vdiag}
\end{equation}
we have the Kraus representation (\ref{ckd}) for the complementary map. For the
diagonal maps $|\psi _{\alpha }\rangle =|e_{\alpha }\rangle ,$ hence from
(\ref{vdiag}) one sees that the diagonal maps are characterized by the property of
having a Kraus representation with simultaneously diagonalizable (i.e. commuting
normal) operators $\tilde{V}_{j}.$ Somewhat more generally, $ \{|\psi _{\alpha
}\rangle\}$ can be an orthonormal base different from $ \{|e_{\alpha }\rangle\}$,
in which case both $\tilde{V}_{k}^*\tilde{V}_{j}$ and
$\tilde{V}_{j}\tilde{V}_{k}^*$ are families of commuting normal operators.

For entanglement-breaking channels the additivity property (\ref{maddi})
(and in fact, (\ref{hsadd}), although not explicitly stated) with arbitrary
second channel was established by Shor \cite{S} and the multiplicativity
property (\ref{n1p}) for all $p>1$ by King \cite{Keb}, using the
Lieb-Thirring inequality. This proof of multiplicativity can be generalized
with almost no changes to the case of entanglement-breaking CP maps. Note
that for diagonal channels (expression (\ref{gdiag}) with $\{|\psi _{\alpha
}\rangle\}=\{|e_{\alpha }\rangle\}$ and $c_{\alpha \alpha }\equiv 1)$ the
properties (\ref{n1p}), (\ref{maddi}) can be established easily because
these channels leave invariant the canonical base in $\mathcal{H}_{\tilde{d}
},$ hence $\nu _{p}(\Phi )=1,\check{H}(\Phi )= 0$ for such channels. Let us
prove for example (\ref{n1p}). (Results for a more general class involving
channels of such kind are given in \cite{F}).

Let $\Phi _{2}$ be an arbitrary CP map, and $\Phi _{1}$ a channel such that $\nu
_{p}(\Phi _{1})=1.$ We have
\begin{equation*}
\nu _{p}(\Phi _{1}\otimes \Phi _{2})=\nu _{p}((\mathrm{Id}_{1}\otimes \Phi
_{2})\circ (\Phi _{1}\otimes \mathrm{Id}_{2}))\leq \nu _{p}(\mathrm{Id}
_{1}\otimes \Phi _{2}),\;
\end{equation*}
where $\mathrm{Id}$ denotes the identity channel. Applying the equality $\nu
_{p}\left( \mathrm{Id}\otimes \Phi \right) =\nu _{p}\left( \Phi \right) $
established in \cite{AHW}, we get
\begin{equation*}
\nu _{p}(\Phi _{1}\otimes \Phi _{2})\leq \nu _{p}(\Phi _{2})=\nu _{p}(\Phi
_{1})\nu _{p}(\Phi _{2}),
\end{equation*}
whence the multiplicativity follows.

However the proof of multiplicativity for diagonal CP maps, that are not
necessarily channels, given in \cite{Kdiag}, is substantially more
complicated (it uses the same method as for the entanglement-breaking maps).
Moreover, this proof seems not to be extendable to the more general class of
CP maps (\ref{gdiag}) where $\left\{ \psi _{\alpha }\right\} $ is not an
orthonormal base, but an arbitrary system of vectors. On the other hand,
theorem 1 implies all the multiplicativity/additivity properties for this
more general class simply by their complementarity to entanglement-breaking
maps and a reference to results in \cite{S,Keb}. Specifically, it implies
the superadditivity property (\ref{hsadd}), which so far was known only for
direct convex sums of the identity and entanglement-breaking channels (e.g.
erasure channel), see \cite{HS}. More precisely, theorem 1 combined with
proposition 3 from \cite{HS} implies property (\ref{hsadd}) for convex
mixtures of either identity or its complementary -- completely depolarizing
channel -- with either entanglement-breaking channel or its complementary.
Therefore additivity of (constrained) $\chi -$capacity holds as well for
such convex mixtures.

4. Let $G$ be a group and $g\rightarrow U_{g}^{A},U_{g}^{B};g\in G;j=1,2,$ be two
(projective) unitary representations of $G$ in $\mathcal{H}_{A},\mathcal{ \ \
H}_{B}$. The CP map $\Phi :\mathfrak{M}(\mathcal{H}_{A})\rightarrow
\mathfrak{M}(\mathcal{H}_{B})$ is \textit{covariant} if
\begin{equation}
\Phi \lbrack U_{g}^{A}\rho U_{g}^{A\ast }]=U_{g}^{B}\Phi \lbrack \rho
]U_{g}^{B}{}^{\ast }  \label{covariant}
\end{equation}
for all $g\in G$ and all $\rho $. The structure of covariant CP maps was
studied in the context of covariant dynamical semigroups, see e. g. \cite{H1}
. In particular, for arbitrary covariant CP map there is the Kraus
representation (\ref{kr}), where $V_{j}$ are the components of a tensor
operator for the group $G$, i. e. satisfy the equations
\begin{equation*}
U_{g}^{B}V_{j}U_{g}^{A\ast }=\sum_{k}d_{jk}(g)V_{k},
\end{equation*}
where $g\rightarrow D(g)=\left[ d_{jk}(g)\right] $ is a matrix unitary
representation of $G.$ It follows that the map complementary to covariant CP
map is again covariant, with $D(g)$ playing the role of the second unitary
representation.

Let us consider in some detail the extreme transpose-depolarizing channel
\begin{equation*}
\Phi (\rho )=\frac{1}{d-1}\left[ I\mathrm{Tr}\rho -\rho ^{T}\right] ,
\end{equation*}
where $\rho ^{T}$ is transpose of $\rho $ in an orthonormal basis $\left\{
e_{j}\right\}$ in $\mathcal{H}=\mathcal{H}_{A}=\mathcal{H}_{B},\dim {\
\mathcal{H}}=d$. This channel breaks the multiplicativity (\ref{n1p}) with $ \Phi
_{1}=\Phi _{2}=\Phi $ for $d>3$ and large enough $p$ \cite{WH}. At the same time it
fulfills the multiplicativity for $1\leq p\leq 2$ \cite{D} and the additivity
(\ref{maddi}), see \cite{MY}, \cite{DHS}. It has the covariance property
\begin{equation*}
\Phi (U\rho U^{\ast })=\bar{U}\Phi (\rho )\bar{U}^{\ast }
\end{equation*}
for arbitrary unitary $U.$ Since
\begin{equation}
\Phi (\rho )=\frac{1}{2(d-1)}\sum_{j,k=1}^{d}\left( |e_{j}\rangle \langle
e_{k}|-|e_{k}\rangle \langle e_{j}|\right) \rho \left( |e_{k}\rangle \langle
e_{j}|-|e_{j}\rangle \langle e_{k}|\right) ,  \label{cas}
\end{equation}
introducing the index $\alpha =(j,k),$ we have the Kraus representation (\ref {kr})
with operators
\begin{equation*}
V_{\alpha }=\frac{1}{\sqrt{2(d-1)}}\left( |e_{j}\rangle \langle
e_{k}|-|e_{k}\rangle \langle e_{j}|\right) .
\end{equation*}
Hence
\begin{eqnarray*}
\tilde{\Phi}(\rho ) &=&\left[ \mathrm{Tr}V_{\alpha }\rho V_{\beta }^{\ast }
\right] _{\alpha ,\beta =\overline{1,d}} \\
&=&\frac{1}{2(d-1)}\left[ \delta _{jj^{\prime }}\langle e_{k}|\rho
|e_{k^{\prime }}\rangle -\delta _{jk^{\prime }}\langle e_{k}|\rho
|e_{j^{\prime }}\rangle -\delta _{kj^{\prime }}\langle e_{j}|\rho
|e_{k^{\prime }}\rangle +\delta _{kk^{\prime }}\langle e_{j}|\rho
|e_{j^{\prime }}\rangle \right].
\end{eqnarray*}
The space $\mathcal{H}_{12}$ in which this matrix acts is tensor product of
two $d-$dimensional coordinate spaces with vectors indexed by $k(k^{\prime
}) $ and $j(j^{\prime }).$ Let $F$ be the operator in $\mathcal{H}_{12}$
which flips the indices $j$ and $k.$ The expression above takes the form
\begin{equation}
\tilde{\Phi}(\rho )=\frac{1}{2(d-1)}(I_{12}-F)(\rho\otimes I_{2} )(I_{12}-F).
\label{ctd}
\end{equation}
This is the complementary channel which shares the
multiplicativity/additivity properties with the channel (\ref{cas}).

By using the decomposition $I_{2}=\sum_{j=1}^{d}|e_{j}\rangle \langle
e_{j}|, $ we have the Kraus representation (\ref{ckd}) for the complementary
channel, where
\begin{eqnarray*}
\tilde{V}_{j}|\psi \rangle &=&\frac{1}{\sqrt{2(d-1)}}(I_{12}-F)( |\psi
\rangle\otimes |e_{j}\rangle ) \\
&=&\frac{1}{\sqrt{2(d-1)}}(|\psi \rangle\otimes |e_{j}\rangle -|e_{j}
\rangle \otimes |\psi\rangle ).
\end{eqnarray*}
The covariance property of the channel (\ref{ctd}) is
\begin{equation*}
\tilde{\Phi}(U\rho U^{\ast })=(U\otimes U)\tilde{\Phi}(\rho )(U^{\ast
}\otimes U^{\ast }),
\end{equation*}
as follows from the fact that $F(U\otimes U)=(U\otimes U)F.$

The case of depolarizing channel
\begin{equation*}
\Phi (\rho )=(1-p)\rho +\frac{p}{d}I\mathrm{Tr}\rho ,\quad 0\leq p\leq \frac{
d^{2}}{d^{2}-1},
\end{equation*}
can be considered along similar lines\footnote{ This case was elaborated jointly
with N. Datta.}. We give only the final result
\begin{equation*}
\tilde{\Phi}(\rho )=S(\rho \otimes I_{2})S,
\end{equation*}
where
\begin{equation*}
S=\sqrt{\frac{p}{d}}I_{12}+\sqrt{d}\left[ -\frac{\sqrt{p}}{d}+\sqrt{ 1-p\left(
\frac{d^{2}-1}{d^{2}}\right) }\left\vert \Omega _{12}\rangle \langle \Omega
_{12}\right\vert \right] ,
\end{equation*}
with $|\Omega _{12}\rangle $ the maximally entangled vector in ${\mathcal{H}}
\otimes {\mathcal{H}}$.

While the depolarizing channel is globally unitarily covariant, the
complementary channel has the covariance property
\begin{equation*}
\tilde{\Phi}[U\rho U^{\ast }]=(U\otimes \bar{U})\tilde{\Phi}[\rho ](U\otimes
\bar{U})^{\ast }
\end{equation*}
for arbitrary unitary operator $U$ in $\mathcal{H}$.

Notice that in both cases the complementary channels have the form
\begin{equation*}
\Phi _{C}(\rho )=S(\rho \otimes I_{B})S^{\ast },
\end{equation*}
where $S:{\mathcal{H}}_{A}\otimes {\mathcal{H}}_{B}\rightarrow {\mathcal{H}}
_{C}$ is such that $\mathrm{Tr}_{{\mathcal{H}}_{B}}S^{\ast }S=I_A.$ There is
a simple general relation between this representation and the second formula
in (\ref{compl}) for arbitrary CP map. Namely, given $V:{\mathcal{H}}
_{A}\rightarrow {\mathcal{H}}_{B}\otimes {\mathcal{H}}_{C}$ choose an
orthonormal basis $\left\{ e_{j}^{\prime}\right\} $ in ${\mathcal{H}}_{B}$
and define $S:{\mathcal{H}}_{A}\otimes {\mathcal{H}}_{B}\rightarrow {\
\mathcal{H}}_{C}$ by the relation $\langle
e_{j}^{\prime}|V=S|e_{j}^{\prime}\rangle ,$ or, more precisely,
\begin{equation*}
\langle \bar{\psi}_{B}\otimes \psi _{C}|V|\psi _{A}\rangle =\langle \psi
_{C}|S|\psi _{A}\otimes \psi _{B}\rangle ,
\end{equation*}
where $\bar{\psi}_{B}$ is complex conjugate in the basis $\left\{ e_{j}\right\} .$
By interchanging the roles of ${\mathcal{H}}_{B},{\mathcal{ \ H }}_{C}$ we of
course obtain a similar representation for the initial map $\Phi _{B}.$ This is in
fact nothing but the dual form (\ref{dual}) of the Stinespring representation, if
$\Phi _{B},\Phi _{C}$ are considered as maps in Heisenberg rather than in
Schr\"{o}dinger picture.

\bigskip The next important class is Bosonic Gaussian channels \cite{HW}.
Any such channel can be described as resulting from a quadratic interaction
with Gaussian environment. It follows that complementary channel is again
Gaussian (see \cite{HW}, Sec. IVB, for an explicit description). As an
example consider attenuation channel with coefficient $k<1$ described by the
transformation
\begin{equation*}
a^{\prime }=ka+\sqrt{1-k^{2}}a_{0}
\end{equation*}
in the Heisenberg picture (to simplify notations we write $a$ instead of $ a\otimes
I_{0}$ and $a_{0}$ instead of $I\otimes a_{0}$), where the mode $ a_{0}$ is in a
Gaussian state. Complementing this transformation with
\begin{equation*}
a_{0}^{\prime }=\sqrt{1-k^{2}}a-ka_{0},
\end{equation*}
we get a canonical (Bogoljubov) transformation implementable by a
Hamiltonian quadratic in $a, a_{0}, a^{\dagger },a_{0}^{\dagger }$. It
follows that the complementary channel is again attenuation channel with the
coefficient $\sqrt{1-k^{2}}$. In the same way, the linear amplifier with
coefficient $k>1$ described by the transformation
\begin{equation*}
a^{\prime }=ka+\sqrt{k^{2}-1}a_{0}^{\dagger },
\end{equation*}
complements to
\begin{equation*}
a_0^{\prime }=\sqrt{k^{2}-1}a^{\dagger }+ka_{0}.
\end{equation*}

More detail on complementary covariant and Gaussian channels will be given
in a subsequent work.

\textbf{Note added in replacement:} Similar ideas, in the context of
channels, are independently developed in the work of C. King, K. Matsumoto,
M. Natanson and M. B. Ruskai \cite{CMNR}.

\bigskip

\textbf{Appendix}

\begin{theorem}
For a CP map $\Phi _{B}:\mathfrak{M}\left( \mathcal{H}_{A}\right)
\rightarrow \mathfrak{M}\left( \mathcal{H}_{B}\right) ,$ there exist a
Hilbert space $\mathcal{H}_{C}$ of dimensionality $d_{C}\leq $ $d_{A}d_{B}$
and an operator $V:\mathcal{H}_{A}\rightarrow \mathcal{H} _{B}\otimes
\mathcal{H}_{C}$, such that the first relation in (\ref{compl}) holds. For
any other such operator $V^{\prime }:\mathcal{H}_{A}\rightarrow \mathcal{H}
_{B}\otimes \mathcal{H}_{C^{\prime }}$ there is a partial isometry $W:
\mathcal{H}_{C}\rightarrow \mathcal{H}_{C^{\prime }}$ such that
\begin{equation}  \label{ww}
V^{\prime }=(I_{B}\otimes W)V,\quad V=(I_{B}\otimes W^{\ast })V^{\prime }.
\end{equation}
\end{theorem}

\textit{Proof.} Consider the algebraic tensor product $\mathcal{L}=\mathcal{ H
}_{A}\otimes \mathfrak{M}(\mathcal{H}_{B})$ generated by the elements $ \psi
\otimes X,\;\psi \in \mathcal{H}_{A},\,X\in \mathfrak{M}(\mathcal{H} _{B}).$ Let us
introduce pre-inner product in $\mathcal{L}$\ with the corresponding square of norm
\begin{equation*}
\Vert \,\sum_{j}\psi _{j}\otimes X_{j}\Vert \,^{2}= \sum_{j,k}\bra{\psi_j} |\Phi^*
(X_{j}^{\ast }X_{k})\ket{\psi_k}=\mathrm{Tr} \sum_{j,k}X_{k}\Phi (
\ket{\psi_k}\bra{\psi_j})X_{j}^{\ast },
\end{equation*}
where $\Phi^*$ is the dual map. This quantity is nonnegative for CP map $ \Phi .$
After factorizing with respect to the subspace $\mathcal{L}_{0}$ of zero norm, we
obtain the Hilbert space $\mathcal{K}=\mathcal{L}/\mathcal{L} _{0}.$ By
construction, $\dim \mathcal{K}\leq d_{A}d_{B}^2.$

Put $V\psi =\psi \otimes I,$ and $\pi (Y)\Psi =\pi (Y)(\psi \otimes X)=\psi
\otimes YX.$ Then $\pi $ is a *-homomorphism $\mathfrak{M}(\mathcal{H}
_{B})\rightarrow \mathfrak{M}(\mathcal{K})$, i. e. a linear map preserving
the algebraic operations and the involution: $\pi (XY)=\pi (X)\pi (Y),\pi
(X^{\ast })=\pi (X)$. Moreover,
\begin{equation}
\bra{\varphi}\Phi^* (X)\ket{\psi}=\braket{\varphi\otimes I} {\psi\otimes X}=
\bra{\varphi}V^{\ast }\pi (X)V\ket{\psi},\qquad X\in \mathfrak{M}(\mathcal{H}
_{B}).  \label{st}
\end{equation}
However any *-homomorphism of the algebra $\mathfrak{M}(\mathcal{H})$ is
unitary equivalent to the ampliation $\pi (X)=X\otimes I_{C}$, where $I_{C}$
is the unit operator in a Hilbert space $\mathcal{H}_{C}$, i.e. we can take $
\mathcal{K}=\mathcal{H}_{B}\otimes \mathcal{H}_{C}$, and (\ref{st}) takes
the form
\begin{equation*}
\bra{\varphi}\Phi^* (X)\ket{\psi}= \bra{\varphi}V^{\ast }\left( X\otimes
I_{C}\right)V\ket{\psi} ,\qquad X\in \mathfrak{M} ( \mathcal{H}_{B}),
\end{equation*}
or
\begin{equation}  \label{dual}
\Phi^*(X)=V^{\ast }\left( X\otimes I_{C}\right)V,
\end{equation}
which is equivalent to the first equation in (\ref{compl}) with $\Phi_B=\Phi$
. It also follows that $\dim \mathcal{H}_C\leq d_{A}d_{B}.$

To prove the second statement, consider the subspace
\begin{equation}
\mathcal{M}=\{(X\otimes I_{C})V\psi :\psi \in \mathcal{H}_{A},X\in \mathfrak{ \
M}(\mathcal{H}_{B})\}\subset \mathcal{K}=\mathcal{H}_{B}\otimes \mathcal{\ H} _{C}.
\end{equation}
It is invariant under multiplication by operators of the form $Y\otimes I_{C} $,
hence it has the form $\mathcal{M}=\mathcal{H}_{B}\otimes \mathcal{ \ M }_{C},
\mathcal{M}_{C}\subset \mathcal{H}_{C}.$ For a minimal representation we should
have $\mathcal{M}_{C}=\mathcal{H}_{C},$ because otherwise there would be a proper
subrepresentation.

Consider a similar subspace $\mathcal{M}^{\prime }=\mathcal{H}_{B^{\prime
}}\otimes \mathcal{M}_{C^{\prime }}$ of the space $\mathcal{K}^{\prime }=
\mathcal{H }_{B}\otimes \mathcal{H}_{C^{\prime }}$ for the second dilation.
Define the operator $R$ from $\mathcal{M}$ to $\mathcal{M}^{\prime }$ by
\begin{equation}
R(X\otimes I_{C})V\psi =(X\otimes I_{C^{\prime }})V^{\prime }\psi .
\label{defW}
\end{equation}
Then $R$ is isometric, since the norms of the vector and of its image under $ R$
are both equal to $\bra{\psi}\Phi^* (X^{\ast}X)\ket{\psi}$ by (\ref{st}). From
(\ref{defW}) we obtain for all $Y\in \mathfrak{M}( \mathcal{H }_{B})$
\begin{equation*}
R(YX\otimes I_{C})V\psi =(Y\otimes I_{C^{\prime }})R(X\otimes
I_{C})V^{\prime }\psi
\end{equation*}
and hence
\begin{equation}
R(Y\otimes I_{C})=(Y\otimes I_{C^{\prime }})R  \label{WYYW}
\end{equation}
on $\mathcal{M}$. Extend $R$ to the whole of $\mathcal{K}$ by letting it
equal to zero on the orthogonal complement to $\mathcal{M}$, then (\ref{WYYW}
) holds on $\mathcal{K}$. Therefore $R=I_{C}\otimes W,$ where $W$
isometrically maps $\mathcal{M}_{C}$ onto $\mathcal{M}_{C^{\prime }}$.
Relation (\ref{defW}) implies (\ref{ww}). $\square $

\textbf{Acknowledgement.} This work was done while the author was the
Leverhulme Visiting Professor at CQC, DAMTP, University of Cambridge. The
author is grateful to Yu. M. Suhov, N. Datta and M. Shirokov for fruitful
discussions.

\end{document}